\documentclass[twoside,12pt]{article}
\usepackage{extsizes}
\usepackage[super,sort&compress,comma]{natbib} 
\usepackage[version=3]{mhchem}
\usepackage[left=1.5cm, right=1.5cm, top=1.785cm, bottom=2.0cm]{geometry}
\usepackage{balance}
\usepackage{mathptmx}
\usepackage{sectsty}
\usepackage{graphicx} 
\usepackage{lastpage}
\usepackage[format=plain,justification=justified,singlelinecheck=false,font={stretch=1.125,small,sf},labelfont=bf,labelsep=space]{caption}
\usepackage{float}
\usepackage{fancyhdr}
\usepackage{fnpos}
\usepackage[english]{babel}
\addto{\captionsenglish}{%
  
}
\usepackage{array}
\usepackage{droidsans}
\usepackage{charter}
\usepackage[T1]{fontenc}
\usepackage[usenames,dvipsnames]{xcolor}
\usepackage{setspace}
\usepackage[compact]{titlesec}
\usepackage{hyperref}

\usepackage{lmodern}
\usepackage{textcomp}

\usepackage{epstopdf}

\definecolor{cream}{RGB}{222,217,201}

\usepackage{bm}

\def\bF{{\bf F}}
\def\bG{{\bf G}}
\def\bGh{{\hat{\bf G}}}
\def\bb{{\bf b}}
\def\br{{\bf r}}

\begin{document}

\pagestyle{fancy}
\thispagestyle{plain}
\fancypagestyle{plain}{
\renewcommand{\headrulewidth}{0pt}
}

\makeFNbottom
\makeatletter
\renewcommand\LARGE{\@setfontsize\LARGE{15pt}{17}}
\renewcommand\Large{\@setfontsize\Large{12pt}{14}}
\renewcommand\large{\@setfontsize\large{10pt}{12}}
\renewcommand\footnotesize{\@setfontsize\footnotesize{7pt}{10}}
\makeatother

\renewcommand{\thefootnote}{\fnsymbol{footnote}}
\renewcommand\footnoterule{\vspace*{1pt}%
\color{cream}\hrule width 3.5in height 0.4pt \color{black}\vspace*{5pt}} 
\setcounter{secnumdepth}{5}

\makeatletter 
\renewcommand\@biblabel[1]{#1}            
\renewcommand\@makefntext[1]%
{\noindent\makebox[0pt][r]{\@thefnmark\,}#1}
\makeatother 
\renewcommand{\figurename}{\small{Fig.}~}
\sectionfont{\sffamily\Large}
\subsectionfont{\normalsize}
\subsubsectionfont{\bf}
\setstretch{1.125} 
\setlength{\skip\footins}{0.8cm}
\setlength{\footnotesep}{0.25cm}
\setlength{\jot}{10pt}
\titlespacing*{\section}{0pt}{4pt}{4pt}
\titlespacing*{\subsection}{0pt}{15pt}{1pt}

\fancyfoot{}
\fancyfoot[LO,RE]{\vspace{-7.1pt}\includegraphics[height=9pt]{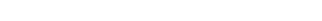}}
\fancyfoot[CO]{\vspace{-7.1pt}\hspace{13.2cm}\includegraphics{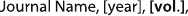}}
\fancyfoot[CE]{\vspace{-7.2pt}\hspace{-14.2cm}\includegraphics{RF}}
\fancyfoot[RO]{\footnotesize{\sffamily{1--\pageref{LastPage} ~\textbar  \hspace{2pt}\thepage}}}
\fancyfoot[LE]{\footnotesize{\sffamily{\thepage~\textbar\hspace{3.45cm} 1--\pageref{LastPage}}}}
\fancyhead{}
\renewcommand{\headrulewidth}{0pt} 
\renewcommand{\footrulewidth}{0pt}
\setlength{\arrayrulewidth}{1pt}
\setlength{\columnsep}{6.5mm}
\setlength\bibsep{1pt}

\makeatletter 
\newlength{\figrulesep} 
\setlength{\figrulesep}{0.5\textfloatsep} 

\newcommand{\topfigrule}{\vspace*{-1pt}%
\noindent{\color{cream}\rule[-\figrulesep]{\columnwidth}{1.5pt}} }

\newcommand{\botfigrule}{\vspace*{-2pt}%
\noindent{\color{cream}\rule[\figrulesep]{\columnwidth}{1.5pt}} }

\newcommand{\dblfigrule}{\vspace*{-1pt}%
\noindent{\color{cream}\rule[-\figrulesep]{\textwidth}{1.5pt}} }

\makeatother


{\includegraphics[height=30pt]{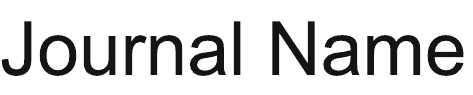}\hfill\raisebox{0pt}[0pt][0pt]{\includegraphics[height=55pt]{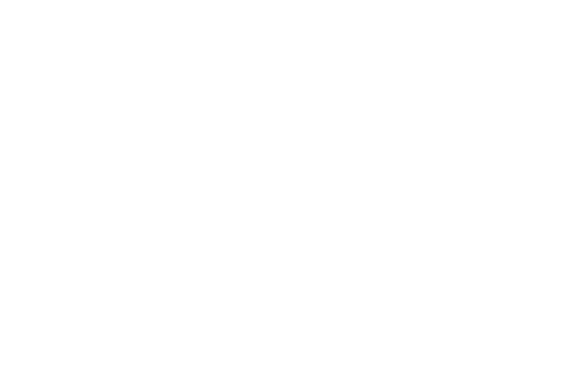}}\\[1ex]
\includegraphics[width=18.5cm]{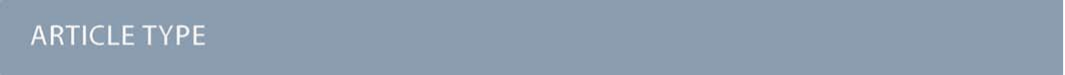}}\par
\vspace{1em}
\sffamily

\noindent\LARGE{\textbf{X-Ray and neutron diffraction patterns of the AlCrTiV high entropy alloy and quaternary Heusler structures}} \\

\noindent\Large{Nedjma Kalliney{$^{a}$} and Michael Widom\textit{$^{a\ddag}$}}~~(\today)\\

\section*{Abstract}
\noindent\normalsize{The quaternary alloy AlCrTiV has been proposed as both a lightweight high entropy alloy and also a functional spin filter material based on the Heusler structure. Experimental investigations to-date, based on X-ray diffraction, offer conflicting interpretations of the structure. Here we simulate diffraction patterns of the various proposed structures to show that neutron diffraction, in particular, can reveal the nature of long-range chemical order and discriminate among distributions of the refractory transition metals. Magnetic contributions to the neutron diffraction are also discussed.}

\vspace{1cm}


\renewcommand*\rmdefault{bch}\normalfont\upshape
\rmfamily
\section*{}
\vspace{-1cm}

\footnotetext{\textit{$^{a}$~Carnegie Mellon University, Department of Physics, Pittsburgh PA, 15213 USA}}
\footnotetext{\textit{$^{\ddag}$~E-mail: widom@cmu.edu}}


\section{Introduction}
\label{sec:Intro}

High entropy alloys~\cite{Yeh04_1,Cantor04} gain configurational entropy from easy chemical substitution among chemically similar elements, for example among the first row of transition metals~\cite{yeh04_2}, or the columns of refractory metals~\cite{Senkov10}. The greatest degree of similarity occurs among neighbors within periodic table rows. Unfortunately, for purposes of structure determination, neighboring elements exhibit low contrast for X-ray diffraction owing to their similar atomic numbers. Neutron scattering lengths take a wide range of magnitudes, and even opposing signs, allowing strong contrast of the first row transition metals among each other and with the heavier refractory metals. In this paper we simulate X-ray and neutron diffraction patterns to illustrate the potential advantages of neutrons for identifying long-range chemical order.

Multicomponent alloys based on the body-centered cubic lattice (BCC, also known at Strukturbericht A2 and Pearson type cI2) can form in a variety of chemically disordered or ordered phases. We will examine the equiatomic quaternary alloy AlCrTiV. It is capable, in principle, of taking a fully disordered BCC structure in which each lattice site has equal probability of occupation by any of the four elements, or it could acquire long-range chemical order of the CsCl prototype (B2, cP2) that consists of two interpenetrating simple cubic sublattices. Certain elements could preferentially occupy cube vertex sites and others prefer the body center. Within a 16-atom $2\times 2\times 2$ supercell, further degrees of ordering could yield structures of prototype NaTl (B32a), BiF$_3$ (DO$_3$), or a variety of Heusler types~\cite{Galanakis_2016,Tian,Hoffman2021} (all sharing Pearson type cF16). The Heusler structures contain four interpenetrating face centered cubic (FCC) sublattices; the four species could each occupy a specific sublattice, or they could be partially mixed in various manners.

\begin{figure}[h!]
  \includegraphics[trim=30mm 4mm 30mm 0mm, clip, width=.3\textwidth]{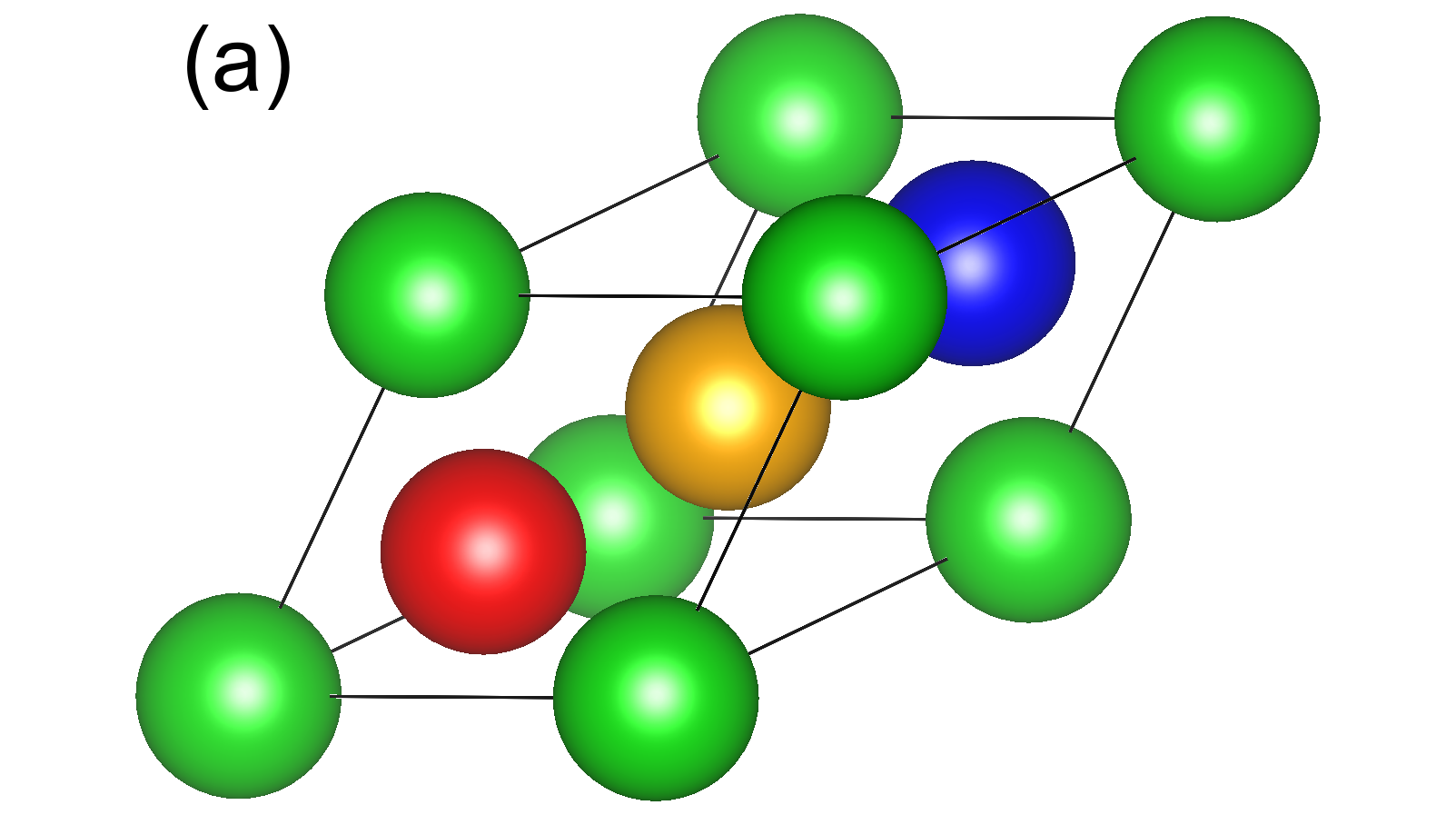}
  \includegraphics[trim=30mm 4mm 30mm 0mm, clip, width=.3\textwidth]{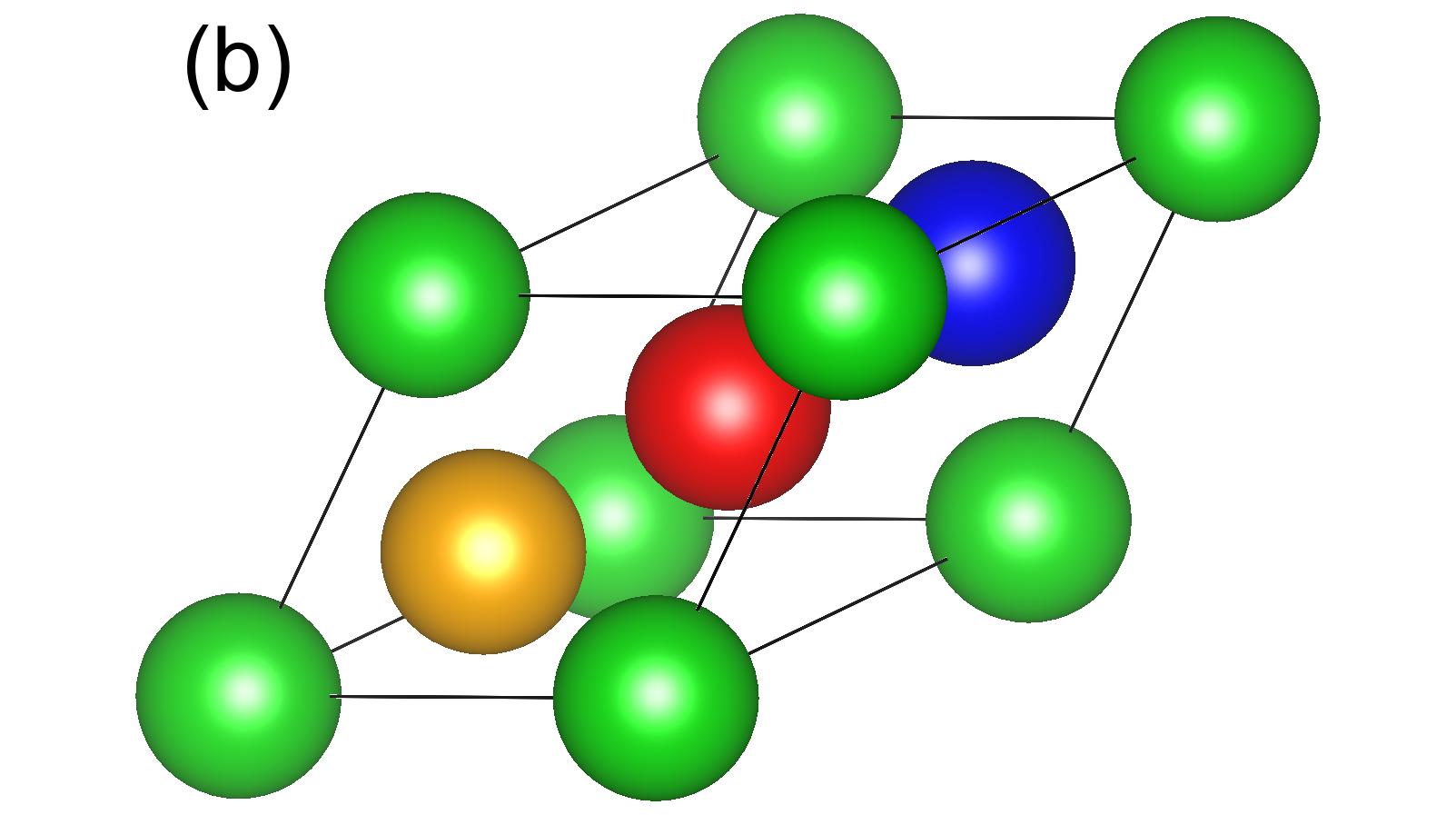}
  \includegraphics[trim=30mm 4mm 30mm 0mm, clip, width=.3\textwidth]{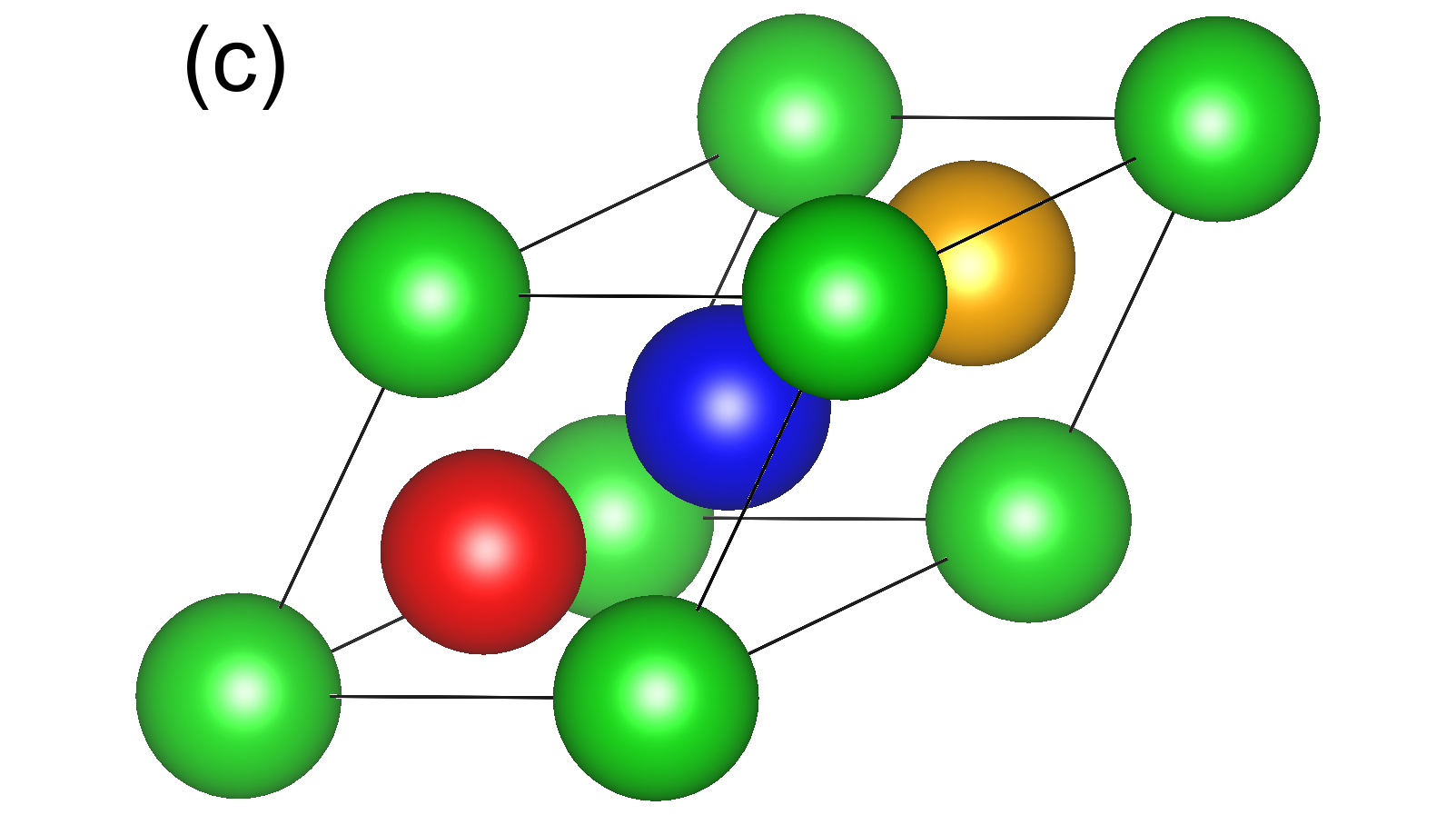}
  \caption{\label{fig:cF16} Primitive cells of the quaternary Heusler structures: (a) Y-I (Al-V-Cr-Ti); (b) Y-II (Al-Cr-V-Ti); (c) Y-III (Al-V-Ti-Cr). Color coding is Al (green); Cr (orange); Ti (blue); V (red).}
\end{figure}

Galanakis, {\em et al.}~\cite{Galanakis_2014} proposed that AlCrTiV would form as a Y-type (prototype LiMgPdSn) quaternary Heusler structure. The four chemical species can be arranged along the diagonal of the cubic cell in three symmetry-independent sequences (see Fig.~\ref{fig:cF16}). We name these types Y-I (diagonal sequence Al-V-Cr-Ti), Y-II (Al-Cr-V-Ti), and Y-III (Al-V-Ti-Cr)~\cite{Widom_2024}. Density functional theory calculations revealed that type III was lowest in energy, and that it was a compensated antiferrimagnetic semiconductor with an extremely high Curie temperature and possible spin-filter applications. Magnetism is driven by antiferromagnetic exchange interactions between Cr and V, which occupy alternating sites of a simple cubic sublattice. Magnetic semiconducting properties were confirmed experimentally~\cite{Stephen2019JAP,Stephen2019PRB}. A DO$_3$ structure (corresponding to Heusler type Y-II with antisite disorder mixing Al with Cr and V) was inferred experimentally but not demonstrated convincingly~\cite{Venkat2018}. Separately, a small fraction of a second phase with L2$_1$ order was reported~\cite{Stephen2019JAP}.

\begin{figure}[h!]
  \includegraphics[trim=30mm 4mm 30mm 0mm, clip, width=.3\textwidth]{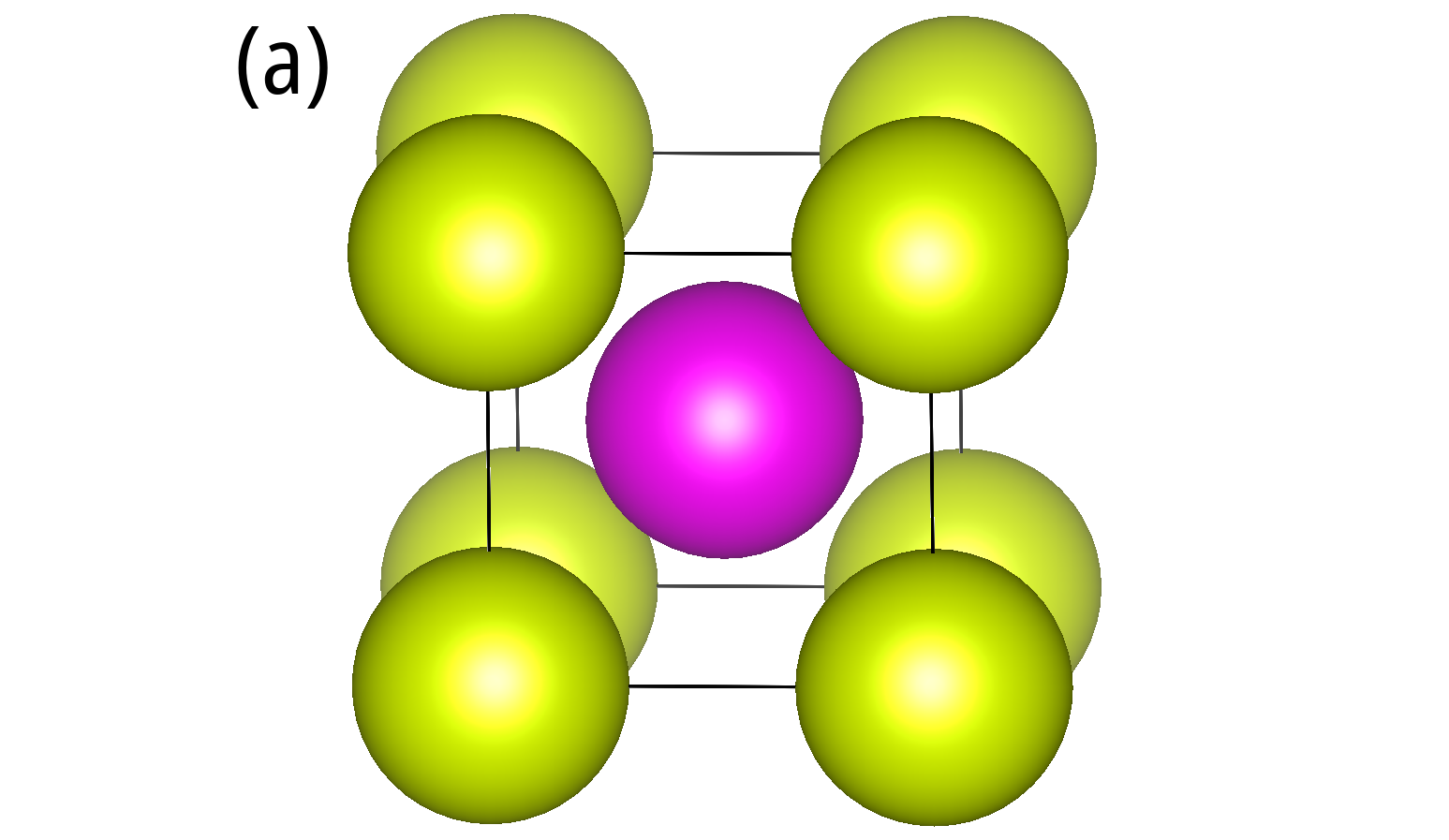}
  \includegraphics[trim=30mm 4mm 30mm 0mm, clip, width=.3\textwidth]{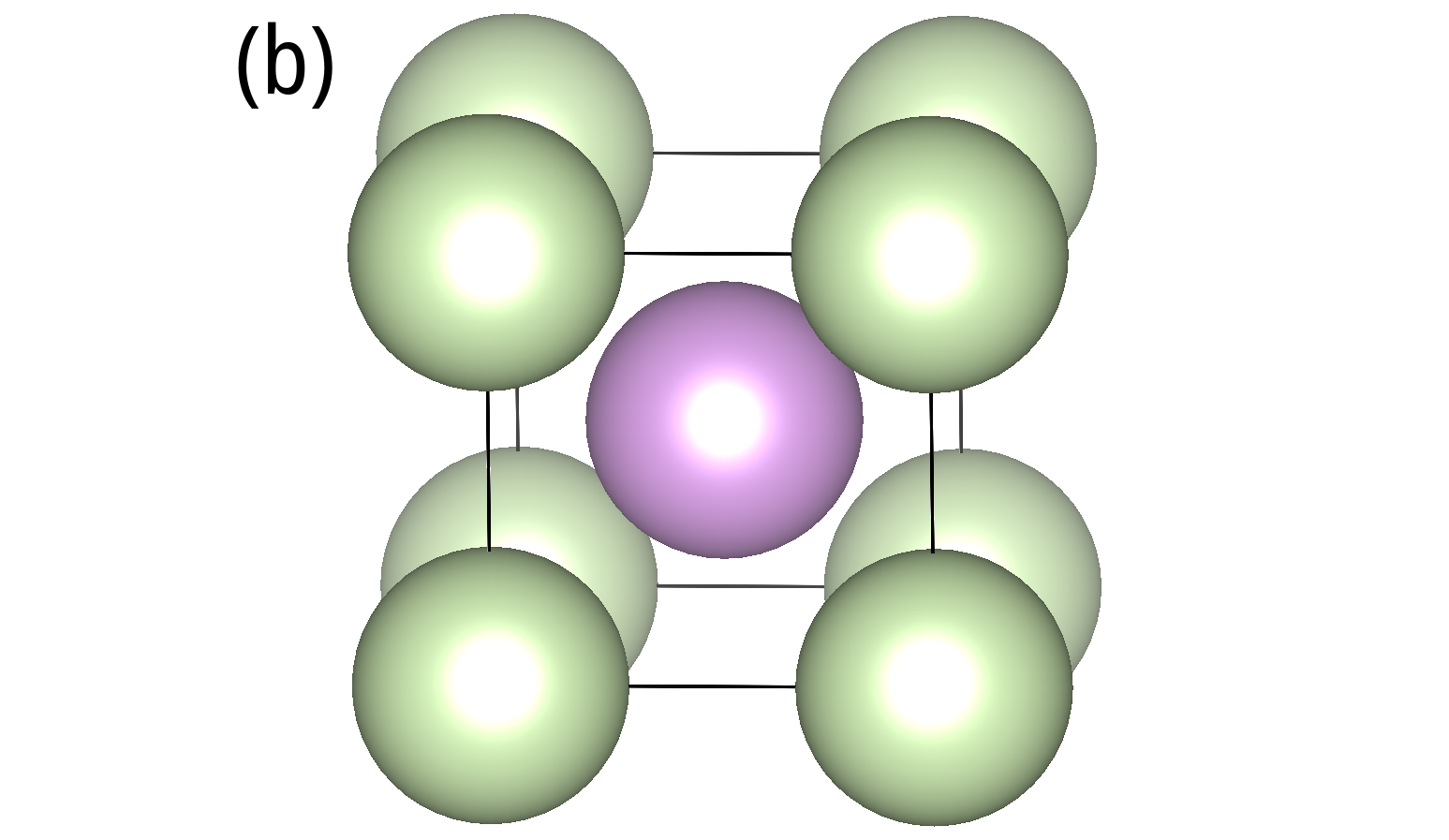}
  \includegraphics[trim=30mm 4mm 30mm 0mm, clip, width=.3\textwidth]{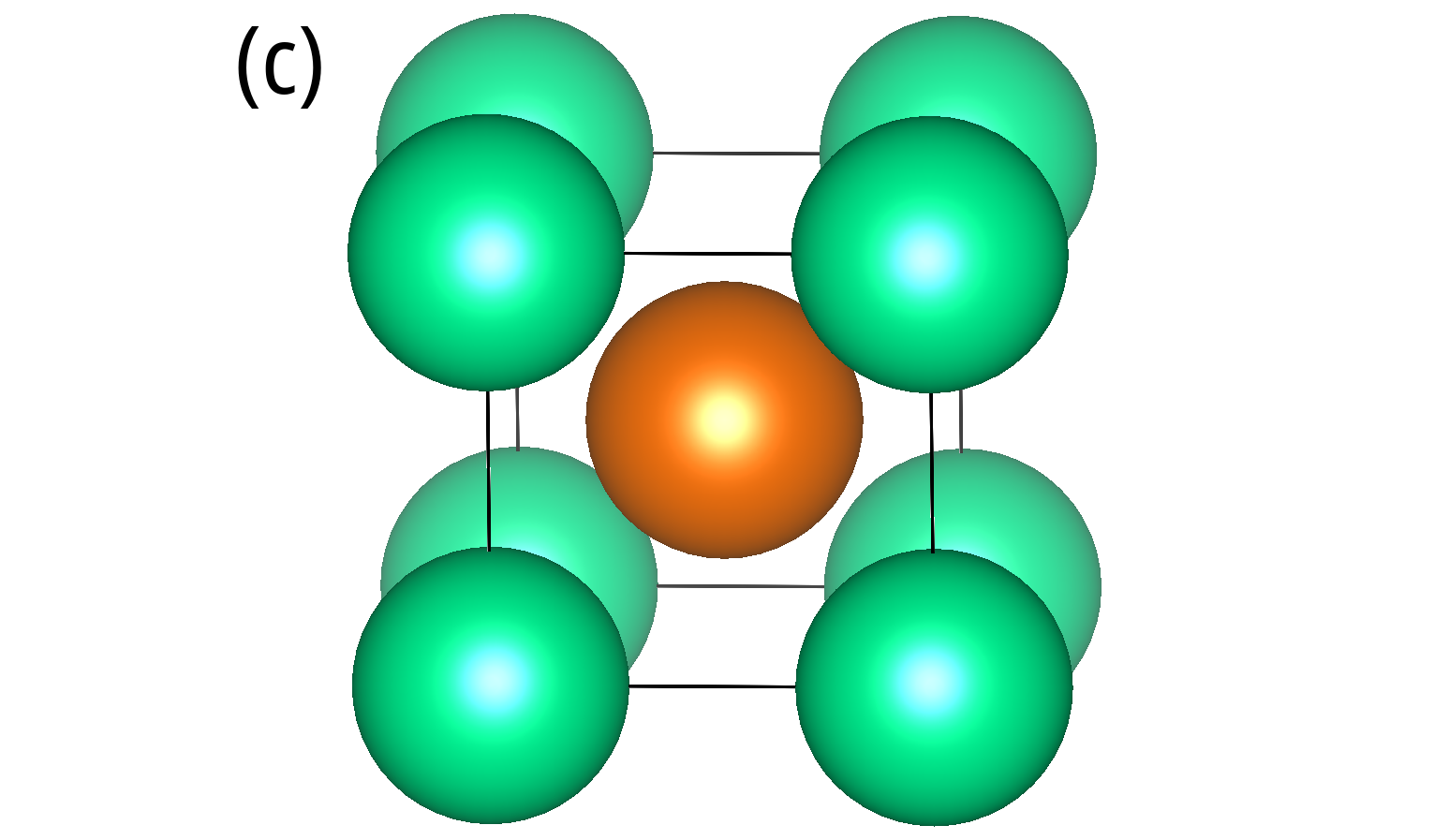}
  \caption{\label{fig:cP2} Primitive cells of the quaternary B2 structures: (a) B2-I-(AlCr)-(TiV); (b) B2-II-(AlV)-(CrTi); (c) B2-III-(AlTi)-(CrV).}
\end{figure}

Independently, Qiu, {\em et al.}~\cite{Qiu2017} proposed that AlCrTiV would form as a single-phase lightweight high entropy alloy. The proposal was confirmed experimentally, and it was shown that the structure was fully disordered BCC at high temperature but that it transformed to partially ordered B2 below 1239K~\cite{Huang2022a}. No indication of Heusler structure was reported. Assuming that two species completely segregate to the cube vertex sites and the other two species occupy the body center sites, there are three variants (see Fig.~\ref{fig:cP2}) that we name as B2-I-(AlCr)-(TiV), B2-II-(AlV)-(CrTi), and B2-III-(AlTi)-(VCr) by analogy with the three quaternary Heusler types. In our notation, elements grouped by parentheses share a common simple cubic sublattice of BCC.

Owing to the similar atomic numbers of the transition metals, and hence their X-ray form factors, X-ray diffraction is not able to discriminate among the differing transition metal orderings. Although X-rays can in principle distinguish Heusler from B2 by observing the (111) peak of the Heusler $2\times 2\times 2$ supercell of the BCC lattice, the intensity is likely to be weak, especially due to antisite disorder of Al with transition metals. For this reason we proposed that neutron diffraction experiments should be performed. The nuclear scattering lengths of Al (3.449), Cr (3.635), Ti (-3.438) and V (-0.3824) lead to excellent contrast among the transition metals. Al and Cr lack scattering length contrast but may be distinguished through magnetic diffraction.

First principles Monte Carlo/molecular dynamics calculations could provide independent evidence on the nature of chemical ordering. In the case of AlCrTiV, simulations using different exchange correlation functionals produce conflicting predictions of chemical order~\cite{Widom_2024}. The PBE~\cite{Perdew96} functional predicts an A2 to B2 transition around 1239K with B2-I-(AlCr)-(TiV) order. The SCAN~\cite{r2SCAN} functional, in contrast, is already B2-I at the melting point, and transforms to Heusler type Y-III below 800K. Identifying the more suitable functional further motivates the neutron experiment.

In the following, we present our methods for calculating the diffraction patterns and the models these will be applied to. We then present our diffraction results in order of increasing complexity, starting from B2 through Heusler and ending with simulated structures.

\section{Methods}
\label{sec:methods}

We wrote a simple powder diffraction calculator to handle X-rays and both nuclear and magnetic neutron diffraction~\cite{github}. Diffraction peaks occur at reciprocal lattice points $\bG=h\bb_1 + k\bb_2 + l\bb_3$, where $\bb_i$ are the primitive reciprocal lattice vectors and $(hkl)$ are Miller indices~\cite{Kittel}. For unpolarized X-ray powder diffraction, the intensity contains three factors~\cite{cullity,vesta}, $I=L\, P\, |F_x|^2$, where $L$ is the Lorentz factor
\begin{equation}
    L = \frac{1}{sin^2(\theta)cos(\theta)},
  \end{equation}
 which accounts for geometrical effects due to powder diffraction, $P$ is the polarization factor
\begin{equation}
    P = \frac{1+cos^2(2\theta)}{2},
\end{equation}
which applies for unpolarized x-rays, and $F_x$ is the structure factor
\begin{equation}
  \label{eq:X-ray}
F_x=\sum_{j} f_je^{-2\pi i \bG\cdot\br_j}.
\end{equation}
Here $f_j$ is the X-ray form factor~\cite{xray_form_factors} of atom $j$ at position $\br_j$. The diffraction angle for reciprocal lattice vector $\bG$ obeys $|\bG|=(4\pi/\lambda)\sin(\theta)$.

The intensity of unpolarized neutron diffraction sums independent nuclear and magnetic contributions,~\cite{rossat-mignod,squires,gsas}, $I = L\, (|F_n|^2+|\bGh\times\bF_m\times\bGh|^2)$, with $L$ the same Lorentz factor as in X-ray diffraction. The nuclear structure factor
\begin{equation}
    F_n = \sum_{j} b_j e^{-2\pi i \bG\cdot\br_j}
\end{equation}
with $b_j$ the neturon scattering length~\cite{neutron_scattering_lengths}. The magnetic structure factor
\begin{equation}
  \bF_m = \sum_{j} \boldsymbol{\mu}_j p_j e^{-2\pi i \bG\cdot\br_j}
\end{equation}
where $\boldsymbol{\mu}_j$ is the magnetic moment of atom $j$ with magnetic form factor~\cite{magnetic_form_factors} $p_j$. In our calculations we employ magnetic form factors appropriate for neutral atoms with quenched orbital moments and we approximate the Land\'e $g$-factors as $g=2$.

For ease of comparison between X-rays and neutrons, and for consistency with typical laboratory equipment, we choose a common wavelength of $\lambda = 1.54059$~\AA~ (Cu-K$\alpha$ radiation) in the calculations that follow. We scale the intensities to the highest peak, and convolute with a Gaussian of width 0.1 to enhance visibility.

Our B2 model structures utilize a 2-atom primitive cubic unit cell, with equal occupation of two species at cube vertex sites(Wyckoff coordinate $1a$ (000)), and equal occupation of the remaining two species at the body center site ($1b$ (\textonehalf \textonehalf \textonehalf)), as illustrated in Fig.~\ref{fig:cP2}. For model Heusler structures we employ the 16-atom $2\times 2\times 2$ supercell of the underlying BCC lattice. The Heusler structure Wyckoff coordinates are (in diagonal order inside the cube): $4a$ (000); $4c$ (\textonequarter \textonequarter \textonequarter); $4b$ (\textonehalf \textonehalf \textonehalf); $4d$ (\textthreequarters\textthreequarters\textthreequarters).

The BCC lattice constant is taken as $a_{\rm BCC}=3.2$~\AA~ so the Heusler lattice constant $a_{\rm Heusler}=6.4$~\AA. Consequently peaks that occur in both BCC and Heusler differ by a factor of $2$ in their Miller indices. In particular, BCC peaks in the standard setting obey $h+k+l=2n$ for integer $n$, while the identical peak indexed in the supercell obeys $h+k+l=4n$. Peaks revealing B2 order obey $h+k+l=2n+1$ using BCC indexing, and $4n+2$ using Heusler indexing. Peaks unique to the Heusler structure have the property that $h, k, l$ are all odd integers.

We have carried out first principles hybrid Monte Carlo/molecular dynamics simulations (FP-MCMD) with replica exchange as described in~\cite{Widom_2024} within 128-atom $4\times 4\times 4$ supercells of BCC using both the PBE~\cite{Perdew96} and the SCAN~\cite{r2SCAN} exchange correlation functionals. A $2\times 2\times 2$ $k$-point grid was applied, using the default plane wave energy cutoff. PBE is a generalized gradient (GGA) functional while SCAN (we use the ``r2'' variant of SCAN) is a meta-GGA that includes a dependence on the electron kinetic energy density. The simulations applied collinear spin polarization and initialized the Cr moments oppositely to Ti and V. We selected 8 independent structures from our simulated ensembles at $T=1000$K and performed short conventional MD at 300K in order to mimic an anneal followed by quench to room temperature. We then map the atomic coordinates to Wyckoff sites within a Heusler-type structure to obtain mean chemical occupancy and magnetic moments.

\section{B2 models}
\label{sec:X-ray}

We begin by illustrating the difficulty of structure discrimination through X-ray diffraction. Table~\ref{tab:B2} gives the chemical species occupations of B2 Wyckoff sites for our three B2 variants. The energies and magnetic moments were determined by randomly distributing the species on appropriate sublattices of a 128-atom $4\times 4\times 4$ supercell of a B2 primitive cell, relaxing at fixed lattice constant, then averaging the moments obtained from density functional theory using the PBE functional.

\begin{table}[h]
  \small
  \centering
  \caption{\ Site occupation and magnetic moments (in $\mu_B$) of B2 models. $\Delta E$ values (in meV/atom) are energies relative to Heusler type-III.}
  \label{tab:B2}
  \begin{tabular*}{0.77\textwidth}{lccccccrcrcrcr}
    \hline
    \multicolumn{5}{c}{Wyckoff} & & \multicolumn{2}{c}{Type I} & &
    \multicolumn{2}{c}{Type II} & &\multicolumn{2}{c}{Type III} \\
    site&x&y&z&occ.& & Chem.&Mom.& & Chem.&Mom.& & Chem.&Mom.\\
    \hline
    1a&0&0&0&0.5& & Al&    0 & & Al&     0& & Al&     0\\
    1a&0&0&0&0.5& & Cr& 1.131& &  V& 2.265& & Ti&-0.292\\
    1b&\textonehalf&\textonehalf&\textonehalf&0.5& & Ti&-0.142& & Cr&-0.123& & Cr& 2.579\\
    1b&\textonehalf&\textonehalf&\textonehalf&0.5& &  V&-1.096& & Ti&-1.344& &  V&-0.457\\
    \hline
    \multicolumn{5}{c}{$\Delta E$} & & \multicolumn{2}{c}{+65} & &
     \multicolumn{2}{c}{+87} & & \multicolumn{2}{c}{+130} \\
    \hline
  \end{tabular*}
\end{table}

\begin{figure}[h!]
\includegraphics[width=.49\textwidth]{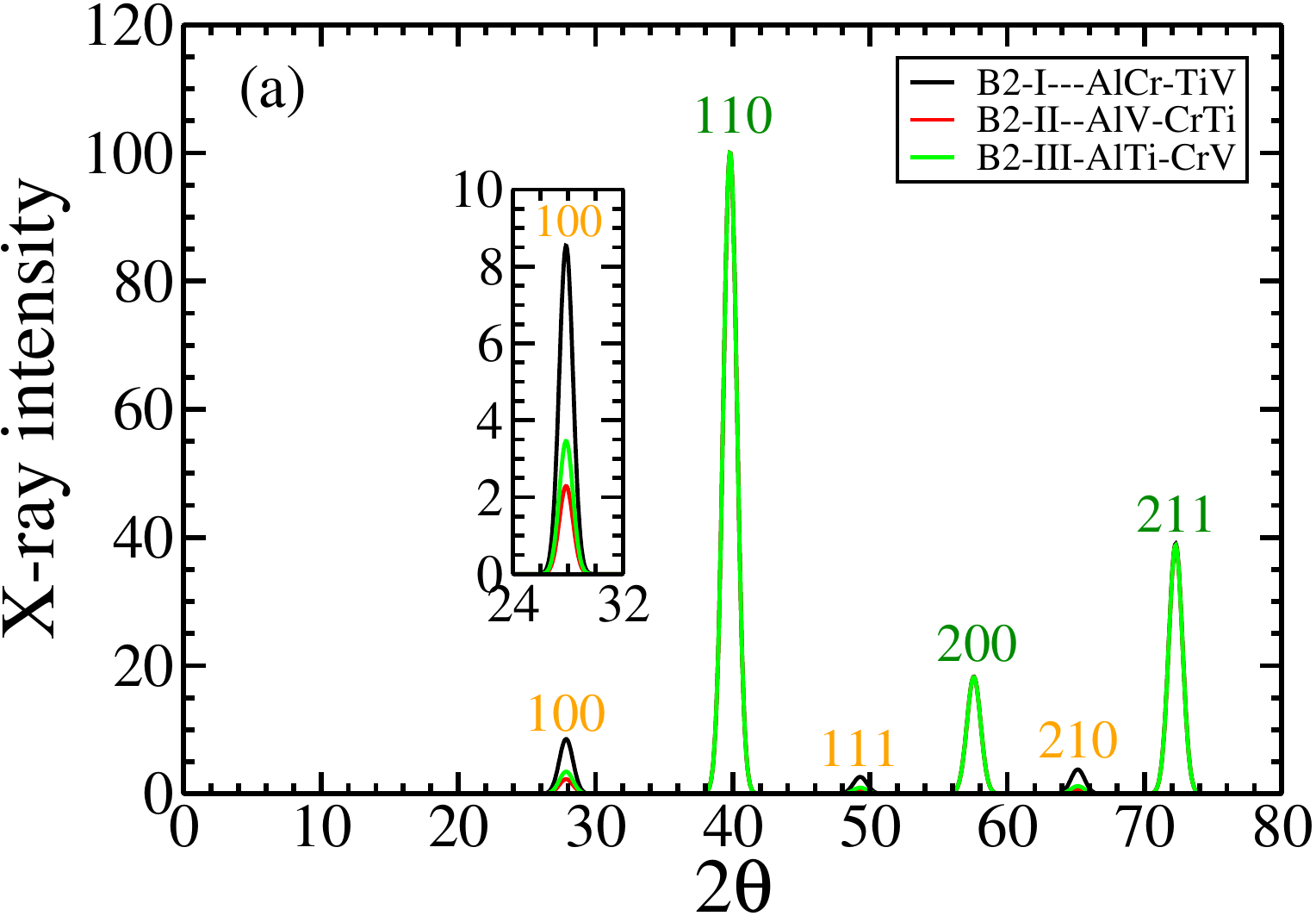}
\includegraphics[width=.49\textwidth]{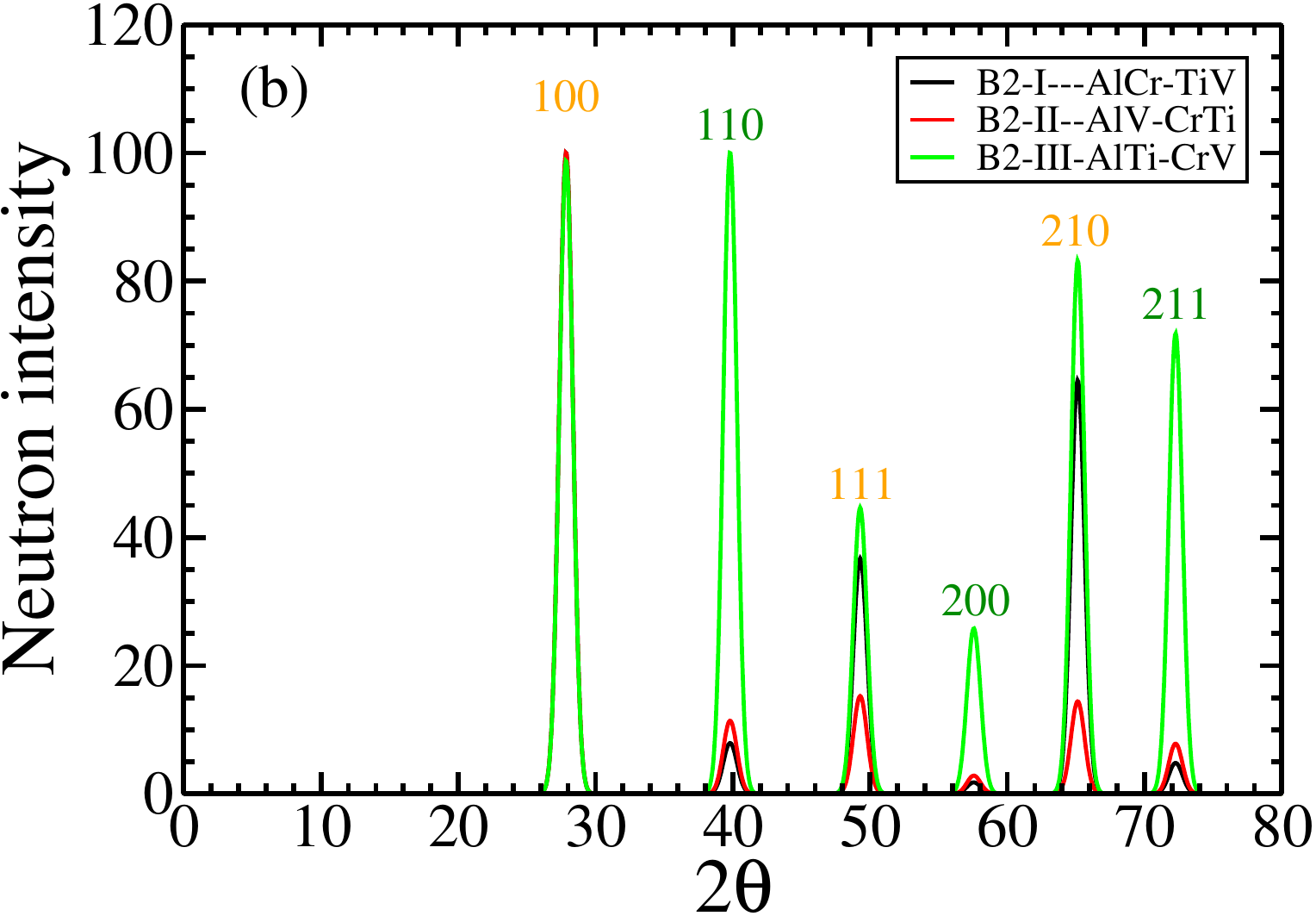}
  \caption{\label{fig:B2} X-Ray (z) and neutron (b) diffraction patterns of  B2 variants. Miller indexing is based on the 2-atom cubic cell of lattice constant 3.2~\AA. Miller indices corresponding to a BCC lattice (A2) are colored in dark green; B2 (CsCl) peak indices are orange Cu-K$\alpha$ radiation ($\lambda=1.5406$~\AA) is assumed for X-rays, and the same wavelength is chosen for the neutron pattern.}
\end{figure}

As shown in Fig.~\ref{fig:B2}a, the predicted diffraction patters of the three B2 variants are nearly indistinguishable from each other. The peaks that obey BCC selection rules (the sum of the Miller indices $h+k+l=2n$) are by far the strongest, but they are insensitive to the distribution of chemical species among the two simple cubic sublattices that comprise the BCC lattice. This is because the exponential factors in Eq.~(\ref{eq:X-ray}) are all identically equal to $1$ for any atom at a BCC lattice site. The peaks that are unique to B2 ($h+k+l=2n+1$) differ according to the patterns of chemical ordering (see inset), but they are all very weak, making it difficult to discriminate among structures with certainty. Distinguishing B2 from BCC depends on the contrast between Al and transition metal form factors, together with strong segregation of Al to a single sublattice of BCC.

Neutron diffraction overcomes the limitations just described for X-rays. The opposite signs of the Ti and V scattering lengths relative to Al and Cr cause the intensities of the ordinary BCC peaks to nearly vanish, while for X-rays these were dominant. Thus the neutron patterns primarily reveal the patterns of chemical ordering, and the B2 (100) peak becomes dominant. The relative intensities of the B2 $(100)$ and $(110)$ peaks easily distinguish B2-I from B2-II and B2-III. This is because in B2-I, Al and Cr occupy a common sublattice of B2; their positive scattering lengths contrast strongly with the negative lengths of Ti and V which occupy the other sublattice.

\section{Heusler models}
\label{sec:neutron}

Table~\ref{tab:Heusler} gives the chemical species occupations for our three Heusler variants. The energies and magnetic moments were obtained from density functional theory using the PBE functional with a high $k$-point density.

\begin{table}[h]
  \small
  \centering
  \caption{\ Site occupation and magnetic moments (in $\mu_B$) of Heusler models. $\Delta E$ values (in meV/atom) are energies relative to Heusler type-III.}
  \label{tab:Heusler}
  \begin{tabular*}{0.77\textwidth}{lcccccrcrcrcr}
    \hline
    \multicolumn{4}{c}{Wyckoff} & & \multicolumn{2}{c}{Type I} & &
    \multicolumn{2}{c}{Type II} & &\multicolumn{2}{c}{Type III} \\
    site&x&y&z& & Chem.&Mom.& & Chem.&Mom.& & Chem.&Mom.\\
    \hline
    1a&0&0&0& & Al&    0 & & Al&     0& & Al&     0\\
    1c&\textonequarter&\textonequarter&\textonequarter& &
                 V&-1.723& & Cr& 2.736& &  V&-2.193\\
    1b&\textonehalf&\textonehalf&\textonehalf& &
                Cr& 1.499& &  V&-1.551& & Ti&-0.509\\
    1d&\textthreequarters&\textthreequarters&\textthreequarters& &
    Ti& 0.330& & Ti&-0.964& & Cr& 2.955\\
    \hline
    \multicolumn{4}{c}{$\Delta E$} & & \multicolumn{2}{c}{+85} & &
     \multicolumn{2}{c}{+30} & & \multicolumn{2}{c}{0} \\
    \hline
  \end{tabular*}
\end{table}

\begin{figure}[h!]
\includegraphics[width=.49\textwidth]{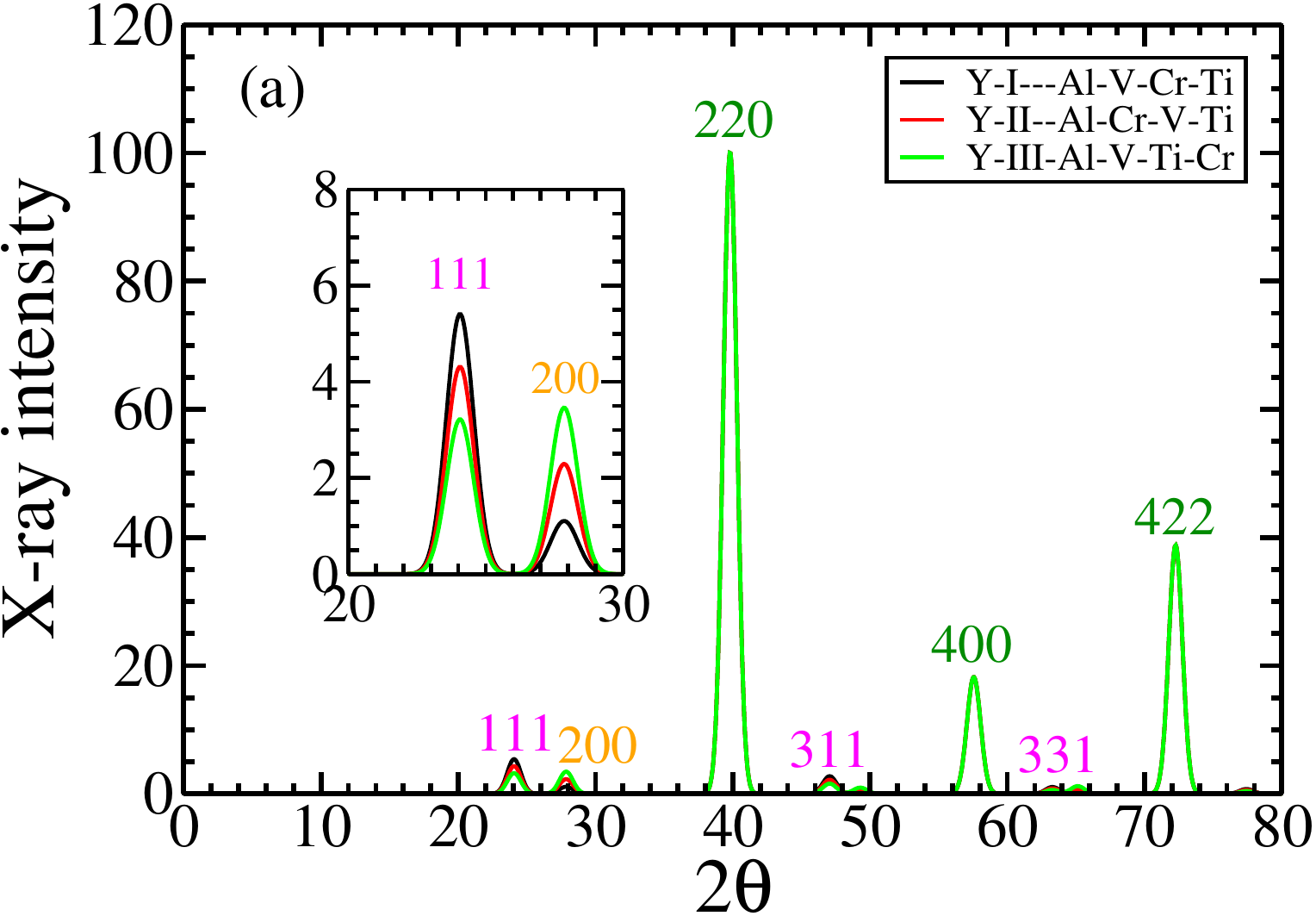}
\includegraphics[width=.49\textwidth]{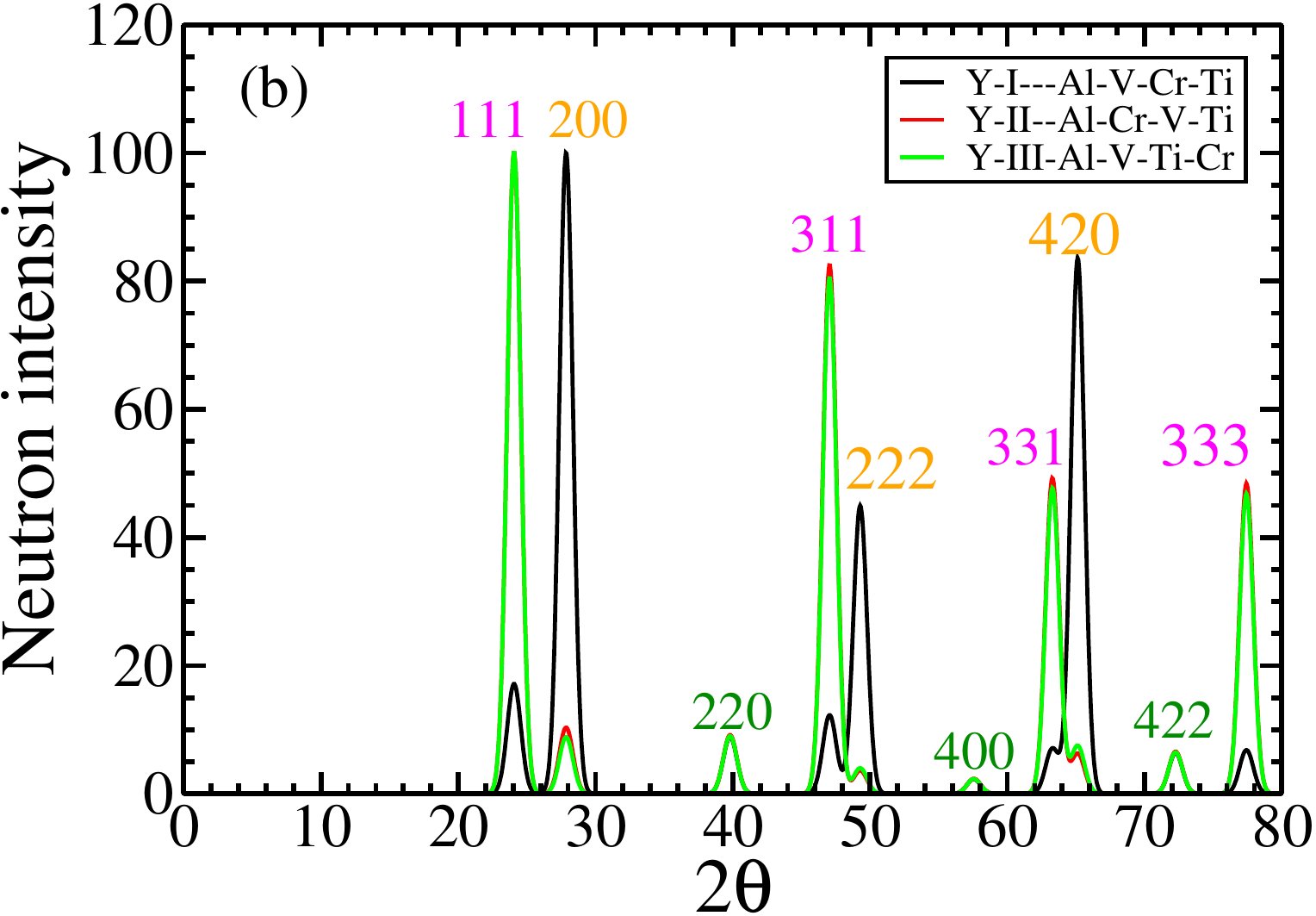}
  \caption{\label{fig:Heusler} X-Ray (a) and neutron (b) diffraction patterns of Heusler Y-type variants. Miller indexing in is based on the 16-atom cubic cell of lattice constant 6.4~\AA. Miller indices corresponding to a BCC (A2) lattice of lattice constant 3.2~\AA~ are colored in dark green; B2 (CsCl) peak indices are orange; peaks unique to Heusler are magenta. Cu-K$\alpha$ radiation ($\lambda=1.5406$~\AA) is assumed for X-rays, and the same wavelength is chosen for the neutron pattern. The inset in (a) enlarges the vicinity of the $(111$) and $(200)$ peaks.}
\end{figure}

The Heusler structures are built upon $2\times 2\times 2$ super cells of BCC or B2, so the peak positions with Miller indices $(hkl)$ in the BCC or B2 cases now double to $(h'k'l')=(2h2k2l)$. Because the Heusler structure has face centered cubic (FCC) symmetry, the indices obey FCC selection rules, $(h'k'l')$ all even or all odd. Thus the Heusler structures exhibit a peak at $(h'k'l')=(111
)$ ({\em i.e.} $(hkl)=(\text{\textonehalf \textonehalf \textonehalf})$). Other than the uniquely Heusler peaks, the diffraction patterns of the Heusler variants are identical to the B2 variants. Observing the $(111)$ peak would be the surest way to distinguish Heusler from B2 or BCC, yet the X-ray intensity of this peak is very weak (see Fig.~\ref{fig:Heusler}) for the same reason that the B2 peak intensity is low - there is very little contrast among the transition metals, so the $(111)$ intensity depends primarily on the Al site. The relative X-ray intensities of the $(111)$ and $(200)$ peaks do contain information that can distinguish among the Heusler types. Their intensity ratio grows from close to 1 for type Y-III, to 2 for type Y-II and over 4 for type Y-I.

As was the case for B2 structures, the neutron diffraction patterns diminish the intensities of the ordinary BCC peaks due to the opposing signs of the different scattering lengths. Again the ratios of $(111)$ to $(200)$ peak intensities provide a clue to the chemical order, but now with much stronger contrast. We find a ratio of 15 for Y-III, 5 for Y-II, and less than 0.4 for Y-I. The B2 $(200)$ peak dominates for Y-I because the positive and negative scattering length elements occupy, respectively, cube vertex (AlCr) and body center (TiV) sites as illustrated in Fig.~\ref{fig:B2} part (a). Sharing Al and Cr equally between vertex and body center sites as in parts (b) and (c) results in near cancelation of the B2 peaks.

\section{Simulated structures}
\label{sec:sim}

We simulated AlCrTiV at temperature of T=1000K as described in Sec.~\ref{sec:methods}. The resulting structures exhibit both short- and long-range order, with Al atoms having low probability to neighbor other Al atoms in either the PBE or SCAN simulations. Ti is the most frequent neighbor of Al in the PBE simulations, while SCAN leads to more frequent Cr neighbors of Al. In order to determine site occupancy statistics, we shift our structures to define the sublattice with the most Al atoms as Wyckoff site $4a$, then perform a reflection (if needed) so that Wyckoff site $4c$ is enriched in Ti (PBE) or in Cr (SCAN). Tables~\ref{tab:PBE} and~\ref{tab:SCAN} give the resulting site occupations and the mean magnetizations of each element on each site. Note the magnetization is partially frustrated, so actual magnitudes and signs vary widely among individual atoms. In general the PBE magnetization is more frustrated, and also the magnitudes are lower than in SCAN.

Inspecting the species distribution from the PBE simulation we see that the occupations of the cube vertex sites (Wyckoff $4a$ and $4b$) are nearly equal, and predominantly Al and Cr. The body center site (Wyckoff $4c$ and $4d$) occupations are also nearly equal, and predominantly Ti and V. The slight deviations from equality are artifacts of the finite system size caused by our shifts and reflections. The observed ordering strongly resembles B2-I-(AlCr)-(TiV), as was proposed in References~\cite{Tian,CA-CPA} on the basis of total energies within the coherent potential approximation.

\begin{table}[h]
\small
  \caption{\ Site occupation and magnetic moments simulated at T=1000K using the PBE functional}
  \label{tab:PBE}
  \begin{tabular*}{0.71\textwidth}{lcccccccccc}
    \hline
    \multicolumn{5}{c}{Occupation} & &\multicolumn{5}{c}{Magnitization} \\
    site&  Al &   Cr  &   Ti  &    V    & &   &    Al &   Cr  &   Ti  &    V \\
    \hline
    4a  & 0.455 & 0.339 & 0.045 & 0.161 & &   &     0 & 1.452 &-0.271 &-0.574 \\
    4c  & 0.053 & 0.156 & 0.527 & 0.263 & &   &     0 & 1.298 &-0.138 &-0.834 \\
    4b  & 0.420 & 0.339 & 0.022 & 0.219 & &   &     0 & 1.114 &-0.273 &-0.697 \\
    4d  & 0.071 & 0.165 & 0.406 & 0.475 & &   &     0 & 0.845 &-0.138 &-0.522 \\
    \hline
  \end{tabular*}
\end{table}

\begin{table}[h]
\small
  \caption{\ Site occupation and magnetic moments simulated at T=1000K using the SCAN functional}
  \label{tab:SCAN}
  \begin{tabular*}{0.71\textwidth}{lcccccccccc}
    \hline
    \multicolumn{5}{c}{Occupation} & &\multicolumn{5}{c}{Magnitization} \\
    site&  Al &   Cr  &   Ti  &    V    & &   &   Al &   Cr  &   Ti  &    V \\
    \hline
    4a  & 0.660 & 0.055 & 0.078 & 0.207 & &   &    0 & 1.990 &-0.809 &-1.280 \\
    4c  & 0.020 & 0.691 & 0.039 & 0.250 & &   &    0 & 2.376 &-0.471 &-1.045 \\
    4b  & 0.289 & 0.105 & 0.246 & 0.359 & &   &    0 & 2.446 &-0.981 &-0.684 \\
    4d  & 0.031 & 0.148 & 0.637 & 0.184 & &   &    0 & 2.121 &-0.987 &-1.280 \\
    \hline
  \end{tabular*}
\end{table}

In contrast to the PBE simulation, the SCAN functional leads to strong segregation into the four separate sublattices, with Al predominantly on $4a$, Cr predominantly on $4c$, and Ti predominantly on $4d$, while V exhibits a weak preference for $4b$. We conclude that the PBE-based structure is best modeled as B2 (at temperature 1000K) while SCAN exhibits a quaternary Heusler structure. The sequence of site preferences under SCAN (Al-Cr-V-Ti) resembles type-II more closely than the energy minimizing type-III (Al-V-Ti-Cr, which is equivalent under inversion to Al-Cr-Ti-V). The energetic favorability of type-III was attributed to the antiferromagnetic interaction of Cr and V at BCC next-neighbor sites. However, we observe that V is broadly distributed among all Wyckoff sites, and in our 1000K Monte Carlo simulations, V has a high acceptance rate (15-20\% compared with $\sim 1$\% for Cr-Ti) with both Cr and Ti, creating substitutional entropy.

\begin{figure}[h!]
\includegraphics[width=.49\textwidth]{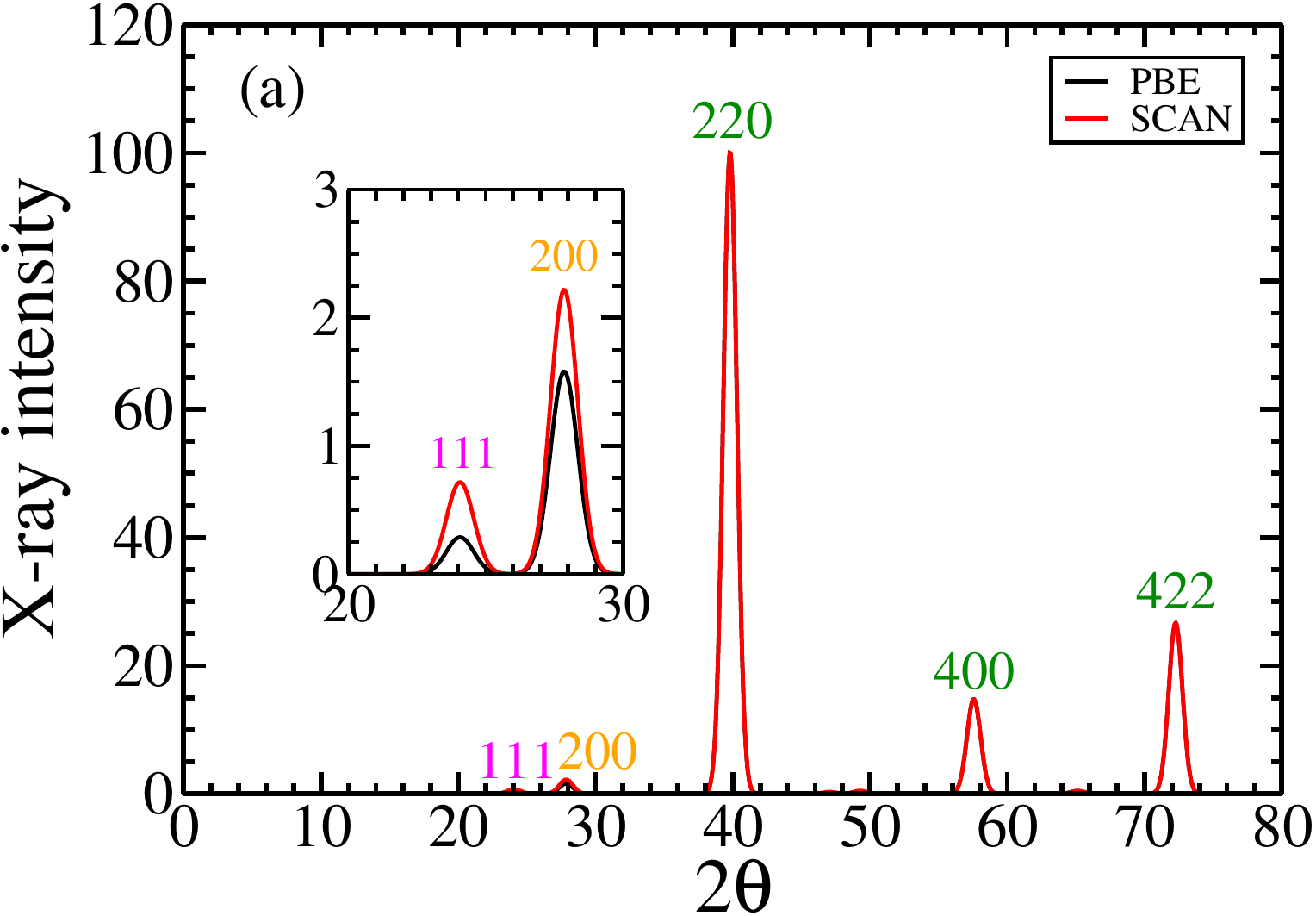}
\includegraphics[width=.49\textwidth]{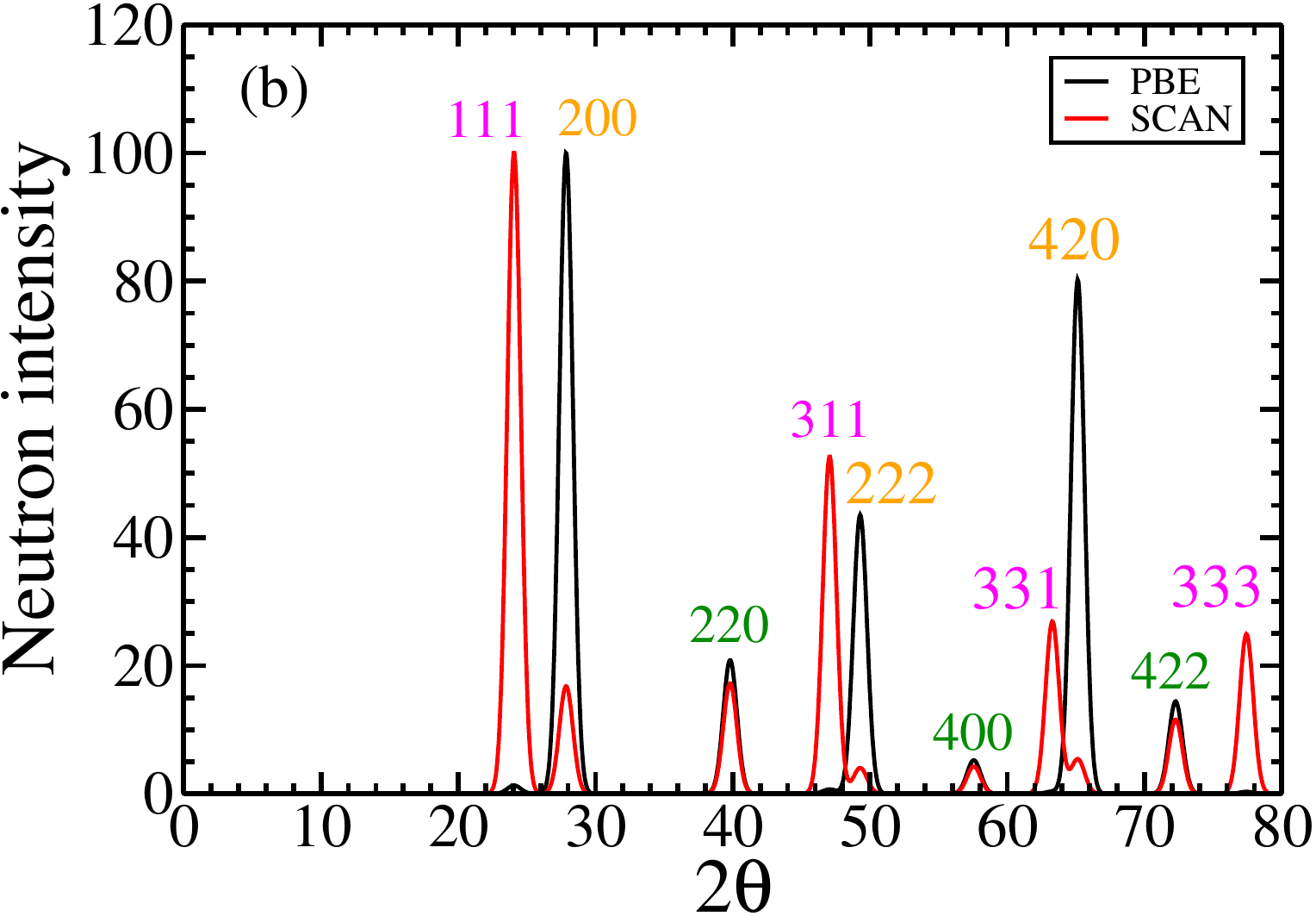}
  \caption{\label{fig:Sim} X-Ray (a) and neutron (b) of structures simulated under PBE and SCAN exchange correlation functionals. Miller indexing in is based on the 16-atom cubic cell of lattice constant 6.4~\AA. Miller indices corresponding to a BCC (A2) lattice of lattice constant 3.2~\AA~ are colored in dark green; B2 (CsCl) peak indices are orange; peaks unique to Heusler are magenta. Cu-K$\alpha$ radiation ($\lambda=1.5406$~\AA) is assumed for X-rays, and the same wavelength is chosen for the neutron pattern. The inset in (a) enlarges the vicinity of the $(111$) and $(200)$ peaks.}
\end{figure}

Figure~\ref{fig:Sim} displays the resulting diffraction patterns. Again, the chemical ordering is revealed by the weak $(111)$ and $(200)$ peaks that are nearly unobservable compared with the strong BCC $(220)$ peak. For the PBE simulation, the neutron simulation exhibits a strong B2 $(200)$ peak. For the SCAN simuation, the neutron pattern exhibits a strong Heusler $(111)$ peak. The $(200)$ peak is also fairly strong, which is more similar to the ideal Heusler type-II pattern than to type-III.

\section{Conclusions}
\label{sec:conclude}
We have shown that X-ray diffraction patterns of AlCrTiV are only weakly sensitive to the difference between BCC, B2 and Heusler order. The sensitivity of X-rays to chemical order rests on the contrasting atomic number of Al relative to the transition metals Ti, V, and Cr, which are neighboring elements in the periodic table with similar X-ray form factors. The strongly contrasting neutron scattering lengths of the transition metals provide far greater sensitivity to the details of chemical order in the B2 cases and especially so in the Heusler cases.

Computer simulations of the chemical order predict outcomes that depend on the choice of exchange-correlation functional. The PBE generalized gradient approximation (GGA) predicted B2-like order at 1000K, while the meta-GGA SCAN predicted Heusler ordering in a $2\times 2\times 2$ supercell of BCC. Our diffraction simulations show that experiments that include neutron diffraction would be uniquely suited to resolving the discrepancy.

So far we have not commented on the magnetic contributions to neutron diffraction. All our calculated neutron patterns include the contributions from unpolarized magnetic diffraction, which is additive to the nuclear component. The magnetic form factors were assumed to be those of neutral atoms, and magnetic moments were obtained through electronic density functional calculations. By adjusting temperature through the anti-ferrimagnetic transition, which lies in the vicinity of $T_c\approx 710$K~\cite{Venkat2018}, it is possible to alter the contrast between the transition metals, which lose magnetism above $T_c$, relative to aluminum, which lacks moments at all temperatures. This can help to distinguish Al from Cr, in particular, as their nuclear form factors are similar.

The calculations presented here examine only the Bragg components of diffraction, reflecting the long-range component of chemical ordering (LRO). Short-range chemical order (SRO) is even more prevalent in HEAs~\cite{Huhn13,Ritchie2021,Woodgate2024}, including in AlCrTiV~\cite{Widom_2024}, and this can lead to diffuse peaks in the diffraction~\cite{Krivoglaz96,Schweika98,Widom16,Ziehl,Foley}. Form factor contrast affects diffuse intensity similarly to Bragg intensity, but other effects including lattice distortion and magnetic frustration contribute additionally. Total scattering investigations could aid in capturing all these contributions~\cite{Billinge_2023,Owen_2024}.

\begin{figure}[h!]
  \centering
\includegraphics[width=.49\textwidth]{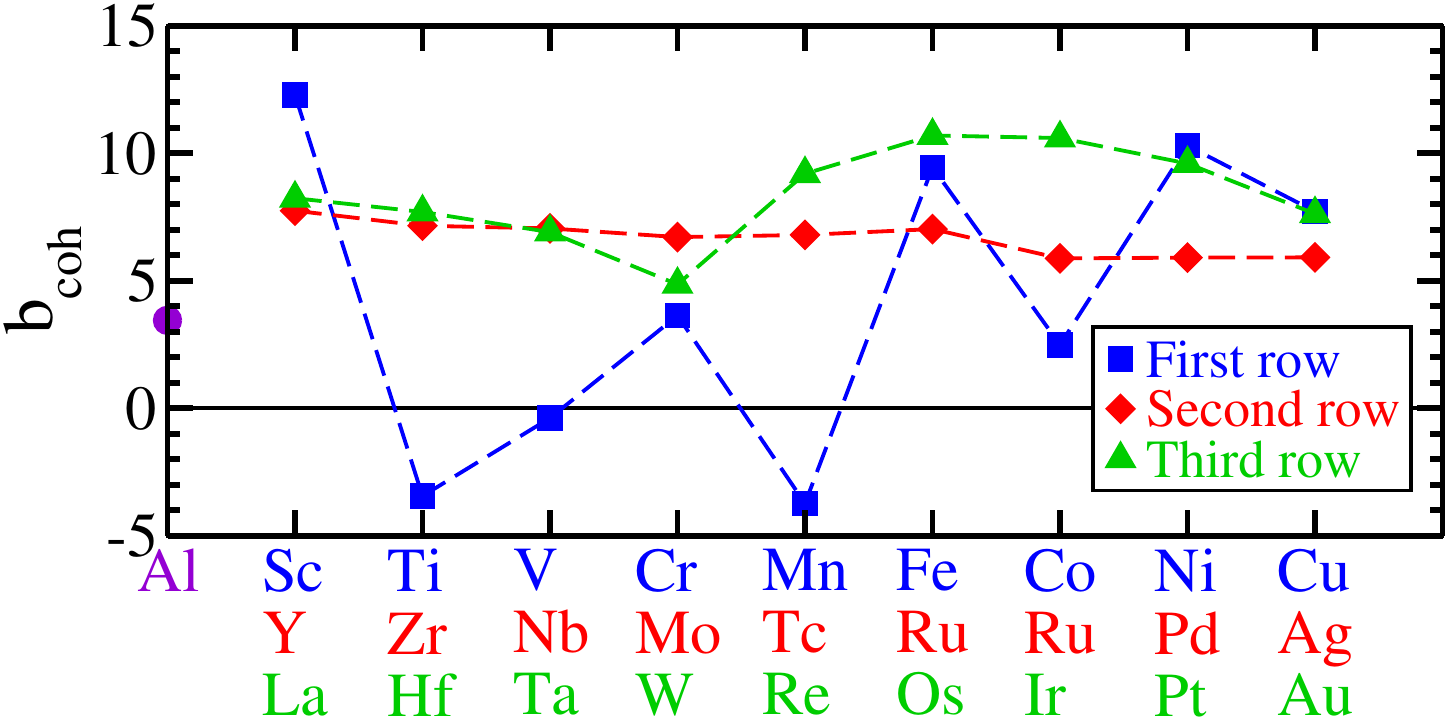}
  \caption{\label{fig:b_coh} Coherent scattering lengths of the transition metals and aluminum.}
\end{figure}

While our discussion centered on the compound AlCrTiV, we emphasize the utility of neutron diffraction experiments for the determination of chemical order in high entropy alloys, owing to the lack of X-ray contrast among many elements that are common in HEAs. Neutron contrast is especially strong within the first row of transition metals~\cite{neutron_scattering_lengths} (see Fig.~\ref{fig:b_coh}). The second row elements have almost no neutron contrast within the row, but they do differ from the first row in both X-ray and in average neutron form factor, especially for the refractory metals. Rare earths (not shown) also offer contrast opportunities, and often can substitute for one another and even for early transition metals.



\section*{Data availability}

Diffraction codes, cif files and resulting diffraction patterns are available at\\
https://github.com/nkalliney1/xray\_neutron\_diffraction\_calculator.

\section*{Acknowledgements}
We thank Nathan Grain, Peter Liaw, Lewis Owen, Alan Goldman and Julia Zaikina for useful discussions. This work was supported by the Department of Energy under Grant No. DE-SC0014506. This research also used the resources of the National Energy Research Scientific Computing Center (NERSC), a US Department of Energy Office of Science User Facility operated under contract number DE-AC02-05CH11231.



\balance


\bibliography{refs} 
\bibliographystyle{rsc} 
\end{document}